\documentclass[a4paper,11pt,twocolumn]{article}

\usepackage[dvips]{graphicx}
\usepackage{amsmath}
\usepackage{mathtext}
\usepackage{cite}

\begin{document}

\title{Four-Probe Methods for Measuring the Resistivity of Samples in the Form of Rectangular Parallelepipeds}

\author{\bf \em L. B. Lugansky $^a$\thanks{e-mail: lugansky@kapitza.ras.ru},
V. I. Tsebro $^b$,\\
{\small \em $^a$Kapitsa Institute for Physical Problems, Russian Academy of Sciences,}\\
{\small \em ul. Kosygina 2, Moscow, 119334 Russia}\\
{\small \em $^b$Lebedev Physical Institute, Russian Academy of Sciences,}\\
{\small \em Leninskii pr. 53, Moscow, 119991 Russia}\\
}

\date{May, 22, 2014}

\maketitle

\begin{abstract}

The problem of measuring the resistivity of isotropic samples of finite dimensions in the form of rectangular parallelepipeds using the four-probe technique was considered. Two variants of contact arrangements were studied: (1) four collinear probes are positioned on one side of a sample symmetrically with respect to the other sides, and (2) two probes on one side of a sample and two on the opposite side are placed precisely in opposite positions and symmetrically with respect to the other sides of the sample (the Schnabel method). Solutions of the problem of the electric field potential distribution in a sample for different positions of the current contacts were found. The solutions were obtained in the form of double series and methods of their summation are presented. The obtained results are extended to the case of measuring the resistivity of anisotropic samples when the resistivity tensor has two independent components. The results of using the developed technique for measuring the resistivity of such a highly anisotropic material as highly oriented pyrolitic graphite using the Schnabel method are presented.

\vspace{36pt}

\end{abstract}

\section{Introduction}

Four-probe methods are widely used to measure the resistivity of materials. The essence of these methods is as follows. Four point contacts are placed on the surface of an investigated sample, a preset current $I$ from an external source is passed through two of them (current contacts), and the potential difference  $\Delta V$ is measured between the two other (potential) contacts. It is obvious that the measured potential difference is proportional to the current value $I$:
\begin{equation}\label{eq:1.1}
\Delta V=RI,
\end{equation}
where the coefficient $R$, which will be called the conditional resistance, is a complex function of the sample geometry, the positions of the current and potential contacts on it, and, certainly, the resistivity $\rho$ of the sample material.

If the studied material is isotropic and is available in the form of long wires or thin long strips, the problem is trivial. The current contacts are placed at the ends of the sample, and the current is then distributed uniformly over the sample cross section $S$ at a sufficient distance from the sample ends. The potential contacts are mounted at a rather long distance from the current contacts at a specified distance $l$ from each other, and the potential difference $\Delta V$ between them is measured at a given current $I$. The resistivity $\rho$ of the material is determined from the formula $R=\Delta V/I=\rho\,l/S$. This technique is well known and widely used in practice.

In more complex cases where the investigated material is represented by, e.g., small crystals (experimental solid state physics), massive semiconductor silicon and germanium wafers (industrial production in microelectronics), massive alloy blocks (metallurgy), etc., the problem becomes much more difficult. To determine $\rho$ from the measurement results, it is necessary to theoretically calculate the form of the function  $R$ in formula (\ref{eq:1.1}) for a particular geometry of the experiment. This is a rather complex problem, especially if the resistivity $\rho$ is anisotropic. However, this problem can be successfully solved for samples with a definite shape at a special arrangement of the measuring electrodes on the sample surface.

The key moment is the solution of the Laplace equation for the electric field potential $u(x,y,z)$ in the investigated material with the corresponding boundary conditions. In the case of an infinite isotropic conducting half-space or an infinite plate, the corresponding solutions are known well and can be found, e.g., in \cite{schnabel64,schnabel67,tikhonov99,pavlov87}.

It usually happens in practice that the studied objects are samples of small dimensions in which the distance between the measuring contacts is comparable to the sample size. In this case, the results described in \cite{schnabel64,schnabel67,tikhonov99,pavlov87} are inapplicable, and a solution of the Laplace equation for a sample with finite dimensions should be sought. This problem can be solved analytically for samples that are shaped as a rectangular parallelepiped \cite{hansen60,mircea63,konkov64,stephen71,polyakov91}. This solution will be presented below in the form that is, in our opinion, more convenient for application than that in \cite{hansen60,mircea63,konkov64,stephen71,polyakov91}. The matter is that the solutions that are obtained using the method of separation of variables have the form of double series, which, as a rule, poorly converge. It is shown below how such series must be correctly summed and correct results can be obtained. The technique we proposed in \cite{lugansky12} allows consideration of samples with finite values of all three dimensions.

In the present paper, we make an attempt to analyze in detail the theoretical questions of such measurements for isotropic substances with the subsequent application of the obtained results to a certain class of anisotropic materials. In connection with this, it should be noted that, e.g., in solid state physics, exactly highly anisotropic systems with layered structures, such as high-temperature superconductors, bismuth-chalcogenide-based topological insulators, highly oriented pyrolitic graphite, etc., attract the greatest interest. The proposed technique with the use of four point contacts is especially convenient for such materials and allows one to obtain the absolute values of the longitudinal (along layers) and transverse (across layers) resistivities and, correspondingly, the value of the resistivity anisotropy.

Two variants of the contact arrangement are considered: (1) four collinear probes are positioned on one side of a sample, and (2) two contacts on one side of a sample and two on the opposite side are placed precisely in opposite positions (the so-called Schnabel geometry \cite{schnabel64,schnabel67}).

\section{Isotropic Sample of Finite Dimensions}

\begin{figure}[h]
\begin{center}
\includegraphics[width=95mm]{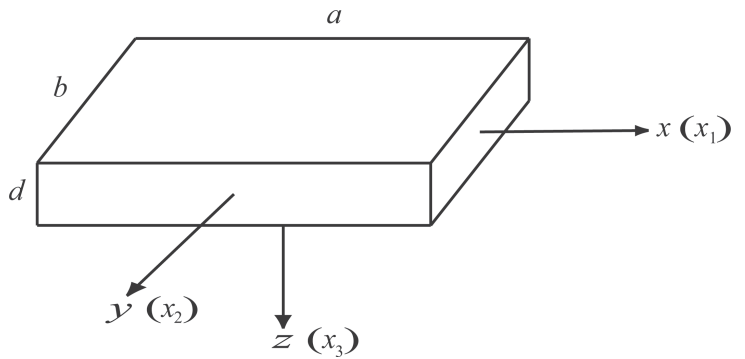}
  \caption{\small Sample of finite dimensions in the form of a rectangular parallelepiped with the length $a$, width $b$, and thickness $d$ and the corresponding directions of the $x$, $y$, and $z$ coordinate axes (an isotropic sample) and $x_1$, $x_2$, and $x_3$ axes (an anisotropic sample). The coordinate origin is at the center of the parallelepiped.}
  \label{fig:fig01}
\end{center}
\end{figure}

Let us consider an isotropic sample in the form of a rectangular parallelepiped with the sides $a$ and $b$ and thickness $d$; the directions of the coordinate axes are shown in Fig.~\ref{fig:fig01}.

\subsection{Two Current Contacts Symmetrically Positioned on One Side of a Sample}

Let us first consider the case where point current contacts are placed on one side of the sample: $z=-d/2$. To simplify this problem, we begin to consider a practical case where both current electrodes are symmetrically positioned on this side (Fig. 2a); i.e., their coordinates are $(\pm x_0, 0, -d/2)$.

\begin{figure}[h]
\begin{center}
\includegraphics[width=60mm]{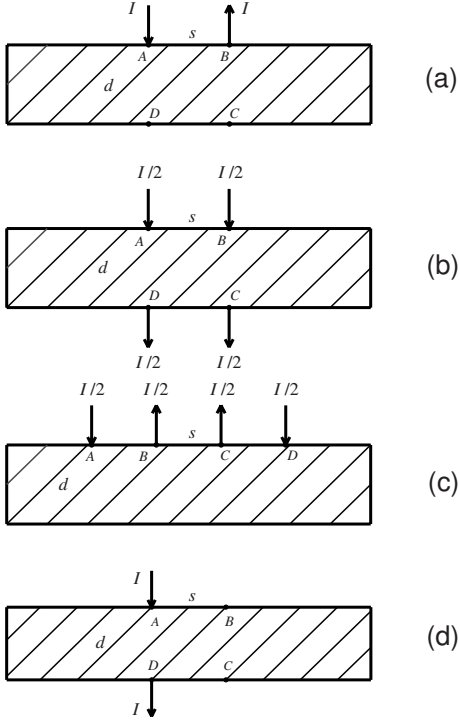}
  \caption{\small Geometry of the arrangements of the contacts on the sides of a sample in the form of a rectangular parallelepiped in the four-probe method: (a) two current contacts $A$, $B$ with the arriving and outgoing current $I$, respectively, are positioned symmetrically on one side of the sample, and if the potential contacts $D$, $C$ are positioned symmetrically with respect to the current contacts on the opposite side, we have the so-called second Schnabel geometry; (b) four current contacts with the entering and outgoing current $I/2$ (as shown with arrows) are positioned on opposite sides of the sample; (c) four current contacts with the entering and outgoing current  $I/2$ (as shown with arrows) are positioned on one side of the sample; and (d) the first Schnabel geometry (the current and potential contacts are, respectively, at the points $A$, $D$ and $B$,  $C$).}
  \label{fig:fig02}
\end{center}
\end{figure}

If a stationary current $I$ flows through the contacts $A$, $B$, an electric field establishes in the sample. Its potential $u(x,y,z)$ satisfies the Laplace equation
\begin{equation}\label{eq:2.1}
u(x,y,z)=\frac{\partial^2u}{\partial x^2}+\frac{\partial^2u}{\partial y^2}+\frac{\partial^2u}{\partial z^2}=0
\end{equation}
with the boundary conditions
\begin{equation}\label{eq:2.2}
    \left\{
    \begin{array}{l} \displaystyle
    \left.\frac{\partial u}{\partial x}\right|_{x=\pm a/2}=0, \left.\frac{\partial u}{\partial y}\right|_{y=\pm b/2}=0,
    \left.\frac{\partial u}{\partial z}\right|_{z= d/2}=0,\\
    \displaystyle
    \left.\frac{\partial u}{\partial z}\right|_{z=-d/2}=-\rho I[\delta(x+x_0)-\delta(x-x_0)]\delta(y).\\
    \end{array} \right.
\end{equation}

The first three conditions in (\ref{eq:2.2}) designate the absence of a current through the interfaces between the sample and environment, and the last condition corresponds to the point nature of the contacts through which the measurement current $I$ flows.

The standard method of separation of variables \cite{tikhonov99} consists in the fact that the solutions of Eq.~(\ref{eq:2.1}) are sought in the form of the product of three functions $u(x,y,z)=X(x) Y(y)Z(z)$, each of which depends only on one variable. The functions  $X(x)$, $Y(y)$, and $Z(z)$ satisfy second-order ordinary differential equations
\begin{equation}
    \frac{d^2W(w)}{dw^2}-cW(w)=0\,. \nonumber
\end{equation}
which, depending on the value and sign of the constant $c$, have one of three fundamental solutions: $W(w)=C_1w+C_2$, $W(w)=A_1\sin\lambda w+B_1\cos\lambda w$, or $W(w)=D_1e^{\lambda w}+D_2e^{-\lambda w}.$

The general solution of Laplace equation (\ref{eq:2.1}) is composed of partial solutions of this kind. Boundary conditions (\ref{eq:2.2}) and the symmetry considerations allow the number of acceptable partial solutions to be significantly reduced. It was shown in \cite{lugansky12} that for the given geometry, the solution of Laplace equation (\ref{eq:2.1}) with boundary conditions (\ref{eq:2.2}) is described by the expression
\begin{eqnarray}\label{eq:2.3}
  &u(x,y,z) \nonumber & \\
   &\displaystyle = -\frac{8\rho Id}{ab}\sum\limits_{k,n=0}^{\infty}\frac{\theta_n\cosh\gamma(z-d/2)}{\gamma d\sinh\gamma d}\sin\frac{(2k+1)\pi x_0}a\nonumber &\\
   &\displaystyle \times \sin\frac{(2k+1)\pi x}a\cos\frac{2n\pi y}b\,,&
\end{eqnarray}
where the quantity $\gamma_{kn}$ is defined by the formula
\begin{equation}\label{eq:2.4}
    \gamma_{kn}=\pi\sqrt{\left(\frac{2k+1}a\right)^2+\left(\frac{2n}b\right)^2}\,,
\end{equation}
and the coefficient $\theta_n$ is
\begin{equation}\label{eq:2.5}
    \theta_n=\left\{ \begin{array}{l}
1,\quad \quad n\ne 0, \\
1/2,\quad n=0.
\end{array} \right.
\end{equation}

This result can be immediately applied to measurements of the resistivity $\rho$ both in the second Schnabel geometry (Fig.~\ref{fig:fig02}a) and in the first collinear geometry, when the outer ($A$ and $D$) and inner ($B$ and $C$) contacts serve as the current and potential contacts, respectively, in a similar way, as is shown in Fig.~\ref{fig:fig02}c for another case. However, this will be done later, and let us now consider two auxiliary problems with four current contacts.

\subsection{Four Current Contacts on Opposite Sides of the Sample}

Let us first consider the auxiliary problem (Fig.~\ref{fig:fig02}b) where there are four symmetrically positioned current contacts; thus, the contacts  $A$ and $B$ have the coordinates ($\pm x_0, 0, -d/2$), and the contacts $C$ and $D$ are positioned at the points ($\pm x_0, 0, d/2$). The current $I/2$ enters each of the contacts ($A$, $B$), and the same current $I/2$ goes out of the opposite contacts ($C$, $D$).

We must solve Laplace equation (\ref{eq:2.1}) with the boundary conditions
\begin{equation*}
    \left\{
    \begin{array}{lc} \displaystyle
    \left.\frac{\partial u}{\partial x}\right|_{x=\pm a/2}=0,\quad \left.\frac{\partial u}{\partial y}\right|_{y=\pm b/2}=0,\\[4mm] \displaystyle
    \left.\frac{\partial u}{\partial z}\right|_{z=\pm d/2}=-\frac{\rho I}2[\delta(x+x_0)+\delta(x-x_0)]\delta(y).\\
    \end{array} \right.
\end{equation*}

From the symmetry considerations, the solution for the potential $u(x,y,z)$ must be symmetrical with respect to the arguments $x$ and $y$.

In this case, the solution of the Laplace equation for the potential  $u(x,y,z)$ has the form \cite{lugansky12}
\begin{eqnarray}\label{eq:2.6}
  &u(x,y,z) \nonumber & \\
  &\displaystyle = -\frac{\rho I}{ab}\,z-\frac{4\rho Id}{ab}{\sum\limits_{k,n=0}^{\infty}}'\frac{\theta_k\theta_n\sinh\gamma z}{\gamma d\cosh(\gamma d/2)} &\\
  &\displaystyle \times
  \cos\frac{2k\pi x_0}a\,\cos\frac{2k\pi x}a \cos\frac{2n\pi y}b\,, &
  \nonumber
\end{eqnarray}
where $\gamma$ is defined by the formula
\begin{equation}\label{eq:2.7}
    \gamma_{kn}=\pi\sqrt{\left(\frac{2k}a\right)^2+\left(\frac{2n}b\right)^2}\,,
\end{equation}
and $\theta_k$ and $\theta_n$ are defined by formula (\ref{eq:2.5}). The prime in the summation sign in formula (\ref{eq:2.6}) means that the term with  $k =  n = 0$ is excluded from the summation.

\subsection{Four Current Contacts on One Side of the Sample}

Let us consider one more auxiliary problem, when there are four current contacts that are positioned symmetrically along a straight line on one side of the sample, as is shown in Fig.~\ref{fig:fig02}c. The contacts $B, C$ and $A, D$ have the coordinates $(\pm x_1, 0, -d/2)$ and $(\pm x_2, 0, -d/2)$, respectively. The current  $I/2$ enters each of the contacts $A$ and $D$, and the same current leaves the contacts $B$ and $C$.

In this case, we must solve Laplace equation (\ref{eq:2.1}) with the boundary conditions
$$
\left\{
\begin{array}{lc} \displaystyle
\left.\frac{\partial u}{\partial x}\right|_{x=\pm a/2}=0, \left.\frac{\partial u}{\partial y}\right|_{y=\pm b/2}=0, \left.\frac{\partial u}{\partial z}\right|_{z=d/2}=0,\\ [4mm] \displaystyle
\left.\frac{\partial u}{\partial z}\right|_{z=-d/2}=-\frac{\rho I}2\,[\delta(x+x_2)-\delta(x+x_1)-\\ \displaystyle \qquad \qquad \qquad -\delta(x-x_1)+\delta(x-x_2)]\delta(y).\\
\end{array} \right.
$$

Assuming for definiteness that the contacts are positioned equidistantly, i.e., $x_1 = s/2$ and  $x_2 = 3s/2$, we obtain the solution for the potential distribution in this problem in the form \cite{lugansky12}
\begin{eqnarray}\label{eq:2.8}
  &u(x,y,z)& \nonumber \\
  & \displaystyle = -\frac{8\rho Id}{ab}{\sum\limits_{k,n=0}^{\infty}}'\frac{\theta_k\theta_n\cosh\gamma (z-d/2)}{\gamma d\sinh(\gamma d/2)}\,\sin\frac{k\pi s}a & \\
  & \displaystyle \times \sin\frac{2k\pi s}a\,cos\frac{2k\pi x}a\,\cos\frac{2n\pi y}b\, ,& \nonumber
\end{eqnarray}
where  $\gamma $ is defined by formula (\ref{eq:2.7}), and and $\theta_k$ and $\theta_n$, by formula (\ref{eq:2.5}).

Now, the obtained results can be applied to practical measuring schemes.

\begin{figure}[h]
\begin{center}
\includegraphics[width=80mm]{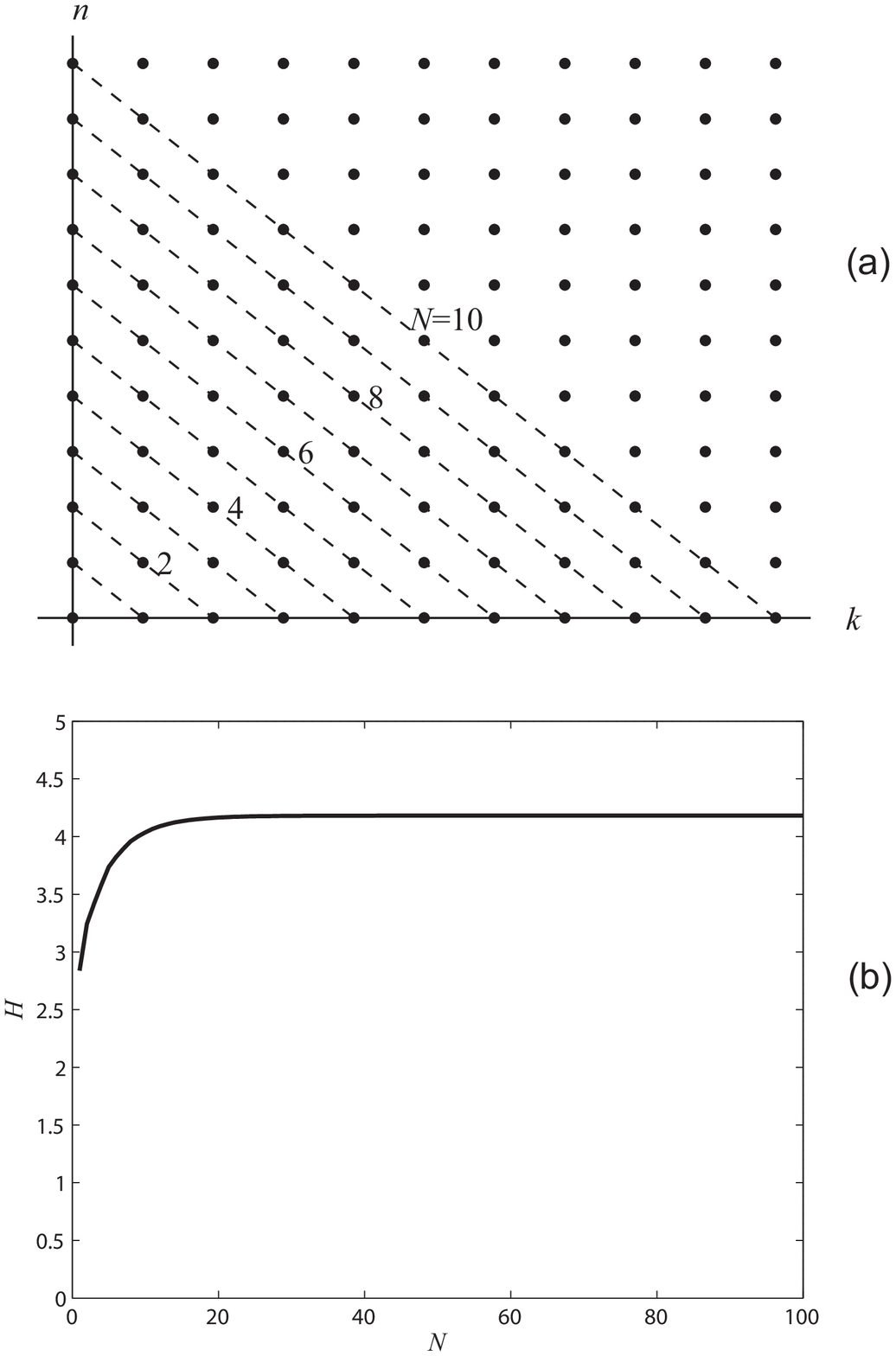}
  \caption{\small  (a) Summation of the double series along the diagonals on the $(k, n)$ plane; (b) the result of calculating the function $H$ as a function of the considered diagonals for a sample with dimensions of  $a$ = 9~mm, $b$ = 3~mm, $d$ = 0.3~mm, and $s$ = 3~mm.}
  \label{fig:fig03}
\end{center}
\end{figure}

\subsection{Measurements using the Schnabel Method. Problem of Series Summation}

Let us first consider the simpler second Schnabel geometry (Fig.~\ref{fig:fig02}a) in which the current contacts have the coordinates $(\pm s/2, 0, -d/2)$ and the potential contacts are placed on the opposite side at the points $(\pm s/2, 0, d/2)$ symmetrically with respect to the current contacts.

Substituting these values of the arguments into formula (4), we easily find the potential difference $V_{DC}$ between the potential contacts:
\begin{equation}\label{eq:2.9}
    V_{DC}=\frac{16\rho Id}{ab}\sum\limits_{k,n=0}^{\infty}\frac{\theta_n\sin^2\frac{(2k+1)\pi s}{2a}}{\gamma d \sinh\gamma d}\ .
\end{equation}

Thus, the conditional resistance  $R_2$ can be expressed by the formula
\begin{equation}\label{eq:2.10}
    R_2=\frac{V_{DC}}I=\frac{\rho}{d}\,H,
\end{equation}
where the function $H$ is defined by the expression
\begin{equation}\label{eq:2.11}
    H=\frac{16 d^2}{ab}\sum\limits_{k,n=0}^{\infty}\frac{\theta_n\sin^2\frac{(2k+1)\pi s}{2a}}{\gamma d\sinh\gamma d}\, ,
\end{equation}
in which the quantity $\gamma_{kn}$ is found using formula (\ref{eq:2.4}).

Hence, the resistivity $\rho $ for an isotropic sample can be found from formula (\ref{eq:2.10}) using the results of one measurement of the resistance $R_2$ in the second Schnabel geometry. As is seen from formula (\ref{eq:2.11}), the function $H$ depends only on the sample dimensions and the distance $s$ between the probes and can be easily calculated using this formula via summation of the double series. This series perfectly converges owing to the presence of exponents in the denominator of its common term (hyperbolic sine). Summing on the plane of the coefficients $(k, n)$ can be performed by different methods: in squares $(k < N, n < N)$ or in circles $(k^2 + n^2 < N^2)$, where the chosen size of the summed area (number $N$) is rather large. However, the most suitable method is to sum along diagonals $(k +  n =  N)$, as shown in Fig.~\ref{fig:fig03}a. The summation procedure terminates at a sufficiently large $N$, when the required accuracy is attained. As a rule, it is sufficient to sum terms on the first several tens of diagonals on the $(k, n)$ plane. Figure~\ref{fig:fig03}b shows, as an example, the change in the result of calculating the function $H$ as a function of the number of considered diagonals $N$ for a sample with dimensions of $a$ = 9~mm, $b$ = 3~mm, $d$ = 0.3~mm, and $s$ = 3~mm.

\begin{figure}[h]
\begin{center}
\includegraphics[width=80mm]{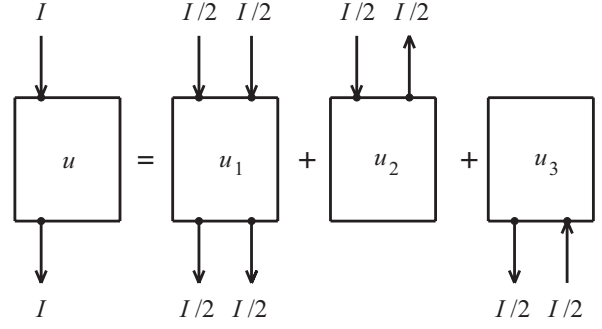}
  \caption{\small  Potential in the first Schnabel geometry as a superposition of three known fields.}
  \label{fig:fig04}
\end{center}
\end{figure}

Let us now consider the problem of the electric field potential distribution for the arrangement of the current and potential probes in accordance with the first Schnabel geometry (Fig.~\ref{fig:fig02}d) in which the current contacts are placed at the points $(-x_0, 0, \pm d/2)$, and the potential contacts are at the points $(x_0, 0, \pm d/2)$. In this case, the sought distribution can be obtained via superposition of three fields in accordance with Fig.~\ref{fig:fig04}. Namely, the desired potential can be represented as the sum $u(x, y, z) =  u_1 (x, y, z) + u_2(x, y, z) + u_3(x, y, z)$, where the potential $u_1$ corresponds to the problem with four current contacts (considered in Subsection 2.2), which are positioned at the same points (Fig.~\ref{fig:fig02}b). The potential $u_2$ corresponds to the second Schnabel geometry (Fig.~\ref{fig:fig02}a) with the measuring current $I/2$, and the potential $u_3$ corresponds to the same second Schnabel geometry with the current $I/2$ and current electrodes positioned on the side $z =  d/2$.

The potential $u_1$ is expressed by formula (\ref{eq:2.6}), and the potential  $u_2$ is expressed by formula (\ref{eq:2.3}) in which $I/2$ must replace the current value $I$. The potential $u_3$ is expressed by the same formula as $u_2$ in which the signs of $x$ and $z$ are reversed, i.e.,  $u_3(x, y, z) =  u_2(-x, y, -z)$. As a result, the following expression for the potential in the sample is obtained:
\begin{equation}\label{eq:2.12}
    \begin{array}{c}\displaystyle
    u(x,y,z) \\[4mm]
    \displaystyle = -\frac{\rho I}{ab}\,z-\frac{4\rho Id}{ab}{\sum\limits_{k,n=0}^{\infty}}'\,\frac{\theta_k\theta_n\sinh\gamma_1 z}{\gamma_1 d\cosh\frac{\gamma_1 d}2}\,\cos\frac{2k\pi x_0}a \\[6mm] \displaystyle \times \cos\frac{2k\pi x}a \cos\frac{2n\pi y}b
    +\frac{4\rho Id}{ab} \\[4mm] \displaystyle \times \sum\limits_{k,n=0}^{\infty}\frac{\theta_n\sinh\gamma_2z}{\gamma_2 d\cosh\frac{\gamma_2 d}2}\,\sin\frac{(2k+1)\pi x_0}a\,\sin\frac{(2k+1)\pi x}a \\ [6mm] \displaystyle \times \cos\frac{2n\pi y}b = -\frac{4\rho Id}{ab}\sum\limits_{k,n=0}^{\infty}\left[\frac{\theta_k\theta_n\sinh\gamma_1 z}{\gamma_1 d\cosh\frac{\gamma_1 d}2}\,\cos\frac{2k\pi x_0}a \right. \\[8mm] \displaystyle \times \cos\frac{2k\pi x}a \cos\frac{2n\pi y}b
    -\frac{\theta_n\sinh\gamma_2z}{\gamma_2 d\cosh\frac{\gamma_2 d}2}\,\sin\frac{(2k+1)\pi x_0}a \\[6mm]
    \displaystyle \left. \times \sin\frac{(2k+1)\pi x}a\cos\frac{2n\pi y}b\right],
    \end{array}
\end{equation}
\vspace{0.1mm}

\noindent keeping in mind that for $k = n = 0$, the first term in the square brackets must be replaced by $z/(4d)$ via a passage to the limit $\gamma_1 \to 0$. Here,
\begin{equation*}
    \begin{array}{c}\displaystyle
    \gamma_1=\pi\sqrt{\left(\frac{2k}a\right)^2+\left(\frac{2n}b\right)^2},\\[6mm]
    \displaystyle \gamma_2=\pi\sqrt{\left(\frac{2k+1}a\right)^2+\left(\frac{2n}b\right)^2}.
    \end{array}
\end{equation*}

Now, we can calculate the potential difference $V_{BC}$ measured in the first Schnabel geometry between the potential contacts $B$ and $C$ (Fig.~\ref{fig:fig02}d). Assuming  $x_0 = s/2$ ($s$ is the distance between the contacts), $x =  s/2$, and $y = 0$, we obtain from (\ref{eq:2.12}):
\begin{equation*}
    \begin{array}{c}\displaystyle
       V_{BC}=\frac{8\rho Id}{ab}
        \sum\limits_{k,n=0}^{\infty}
        \left[\frac{\theta_k\theta_n}{\gamma_1d}\tanh\frac{\gamma_1d}2\,\cos^2\frac{k\pi s}a \right. \\[6mm]
        \displaystyle \left. -\frac{\theta_n}{\gamma_2d}\tanh\frac{\gamma_2d}2\,\sin^2\frac{(2k+1)\pi s}{2a} \right]
     \end{array}
\end{equation*}

Thus, the conditional resistance $R_1$ measured in the first Schnabel geometry is determined as
\begin{equation}\label{eq:2.13}
    R_1 = \frac{V_{BC}}{I} = \frac{\rho}{d}\,G
\end{equation}
where the function $G$ has the form
\begin{equation}\label{eq:2.14}
    \begin{array}{c}\displaystyle
      G=\frac{8d^2}{ab}
    \sum\limits_{k,n=0}^{\infty}
    \left[\frac{\theta_k\theta_n}{\gamma_1d}\tanh\frac{\gamma_1d}2\,\cos^2\frac{k\pi s}a \right.\\[6mm]
      \displaystyle \left.
      -\frac{\theta_n}{\gamma_2d}\tanh\frac{\gamma_2d}2\,\sin^2\frac{(2k+1)\pi s}{2a}
    \right].
    \end{array}
\end{equation}

Unfortunately, it is impossible to use formula (\ref{eq:2.14}) because the double series in it converges poorly, and for the potential contacts in this geometry, which are positioned at the points $(s/2, 0, \pm d/2)$ on the sample surface, the summation result does not correspond to the physical meaning of the problem. This is due to the fact that, in this case, the observation points correspond to places where the fictitious current contacts are located in accordance with Fig.~\ref{fig:fig04}.

In order to get out of this difficult situation, we shall consider hypothetic potential contacts that, in contrast to real contacts, are located not at the sample surface but inside it at the points $(s/2, 0, \pm(d/2-\varepsilon) )$, where $\varepsilon/d$ is a rather small value. The result we need for the contacts that are on the sample surface can be obtained at $\varepsilon$ tending to zero.

Adopting this approach, we obtain the following expression for the measured potential difference $V_{BC}$ from formula (\ref{eq:2.12}):
\begin{equation*}
    \begin{array}{c}
       \displaystyle V_{BC}=\frac{8\rho Id}{ab}
    \sum\limits_{k,n=0}^{\infty}
    \left[\frac{\theta_k\theta_n}{\gamma_1d}\,\frac{\sinh\gamma_1(\frac d2-\varepsilon)}{\cosh\frac{\gamma_1d}2}\cos^2\frac{k\pi s}a \right.\\[6mm]
    \displaystyle \left. -\frac{\theta_n}{\gamma_2d}\,\frac{\sinh\gamma_2(\frac d2-\varepsilon)}{\cosh\frac{\gamma_2d}2}\sin^2\frac{(2k+1)\pi s}{2a}
    \right].
     \end{array}
\end{equation*}

From here, as was above, we obtain the expression for the function $G$, which is written in a more suitable form for practical calculations, by separating the term corresponding to  $k = n = 0$ and expressing the hyperbolic sine and cosine through the exponents:
\begin{equation}\label{eq:2.15}
    \begin{array}{c}\displaystyle
    G=\frac{8\pi d^2}{ab}\left\{\frac 18-\frac{\varepsilon}{4d}-\frac a{2\pi d}\,\frac{\sinh\frac{\pi}a(\frac d2-\varepsilon)}{\cosh\frac{\pi d}{2a}}\sin^2\frac{\pi s}{2a} \right.\\[6mm]

    +\left.{\sum\limits_{k,n=0}^{\infty}}'\displaystyle
    \left[\frac{\theta_k\theta_n}{\gamma_1d}\,\frac{e^{-\gamma_1\varepsilon}-e^{-\gamma_1(d-\varepsilon)}}{1+e^{-\gamma_1d}}\,
    \cos^2\frac{k\pi s}a \right.\right. \\[6mm]

    \displaystyle \left. \left. - \frac{\theta_n}{\gamma_2d}\,\frac{e^{-\gamma_2\varepsilon}-e^{-\gamma_2(d-\varepsilon)}}{1+e^{-\gamma_2d}}\,
    \sin^2\frac{(2k+1)\pi s}{2a}\right]\right\}. \\
    \end{array}
\end{equation}

\vspace{4mm}

\begin{figure}[h]
\begin{center}
\includegraphics[width=80mm]{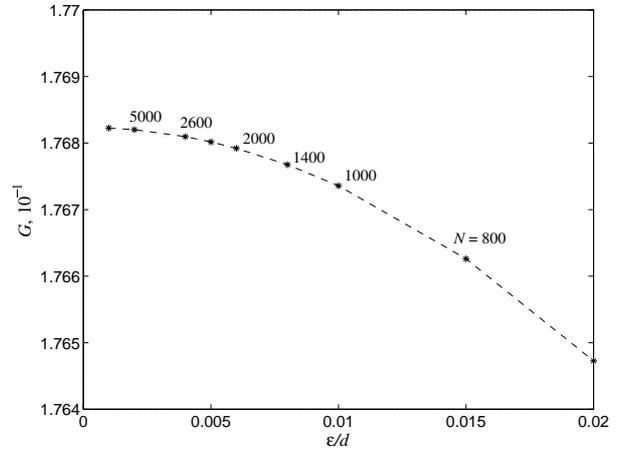}
  \caption{\small  Dependence of the function $G$ on the depth of embedding  $\varepsilon$ of the potential contacts. Numbers near the curve denote the approximate number of diagonals $N$ on the $(k, n)$ plane.}
  \label{fig:fig05}
\end{center}
\end{figure}

The double series in (\ref{eq:2.15}) converges because of the presence of the corresponding exponents, and the convergence rate increases with an increase in the deepening $\varepsilon$. However, to obtain the sufficient accuracy, one has to consider several hundred or even thousand diagonals on the $(k, n)$ plane.

As an example, Fig.~\ref{fig:fig05} shows the plot that presents the results of calculating the function $G$ from formula (\ref{eq:2.15}) for a sample with dimensions of  $a = 9$~mm, $b = d = s = 3$~mm at different $\varepsilon$ values. The approximate number of diagonals $N$ on the plane of indices $(k, n)$, over which double series (\ref{eq:2.15}) must be summed for the required accuracy to be obtained, is also indicated in this curve. This curve is a parabola from which it is easy to find the value of $G$ at  $\varepsilon = 0$ by extrapolation. If two results  $G_1 (\varepsilon_1)$ and $G_2 (\varepsilon_2)$ are taken and the curve  $G(\varepsilon)$ is approximated by the parabola $G(\varepsilon)=G(0)+g\varepsilon^2$, then we obtain the following value for $G(0)$:
\begin{equation}\label{eq:2.16}
    G(0)=\frac{G_1\varepsilon_2^2-G_2\varepsilon_1^2}{\varepsilon_2^2-\varepsilon_1^2}.
\end{equation}

If $\varepsilon_2=2\varepsilon_1$, formula (\ref{eq:2.16}) transforms into $G=(4G_1-G_2)/3$, and the result very weakly depends on $\varepsilon_1$. Just in this way, using two $G$ values, the function $G(0)$ was further calculated.

\subsection{Four Collinear Contacts on One Side of the Sample}

Let us now consider a widely used practical case where all the four probes are positioned on one side of a sample equidistantly and symmetrically along one line (e.g., as is shown in Fig.~\ref{fig:fig02}c). We define the first collinear geometry as the configuration of probes when the current contacts are at the points $(A,D)$ and have the coordinates $(\pm 3s/2,0,-d/2)$, and the potential contacts $(B,C)$ have the coordinates $(\pm s/2,0,-d/2)$, where  $s$ is the distance between the neighboring probes.

Using formula (\ref{eq:2.3}), we find the potential difference $V_{BC}$ and the conditional resistance
\begin{equation*}
    \begin{array}{c}\displaystyle
       R_1=\frac{V_{BC}}I=\frac{16\rho d}{ab}\sum\limits_{k,n=0}^{\infty}\frac{\theta_n\,\cosh\gamma d}{\gamma d\,\sinh\gamma d}\,\sin\varphi_k\,\sin 3\varphi_k ,\\[6mm]
       \displaystyle \varphi_k=\frac{(2k+1)\pi s}{2a}
     \end{array}
\end{equation*}

Thus, by analogy to formulas (\ref{eq:2.10}) and (\ref{eq:2.13}), $R_1$ can be written in the form
\begin{equation*}
    R_1 = \frac{\rho}{d}\,L_1
\end{equation*}
where
\begin{equation}\label{eq:2.17}
    L_1=\frac{16 d^2}{ab}\sum\limits_{k,n=0}^{\infty}\frac{\theta_n\,\cosh\gamma d}{\gamma d\,\sinh\gamma d}\,\sin\varphi_k\,\sin 3\varphi_k,
\end{equation}
and the parameter $\gamma_{kn}$ is determined by formula (\ref{eq:2.4}). By measuring  $R_1$ and calculating the value of the function $L_1$ using formula (\ref{eq:2.17}), we can find the material resistivity $\rho$.

When the function $L_1$ is calculated, we also encounter a very poor convergence of the double series in formula (\ref{eq:2.17}). When summing in squares $(k \leq N, n \leq N)$, a partial sum of this series oscillates around a certain average value without damping with an increase in $N$.
To obtain a correct result, the summation should be performed along diagonals $(k +  n =  N)$ on the $(k, n)$ plane. In this case, these oscillations gradually decay with an increase in $N$ and approach the average value.

\begin{figure}[h]
\begin{center}
\includegraphics[width=80mm]{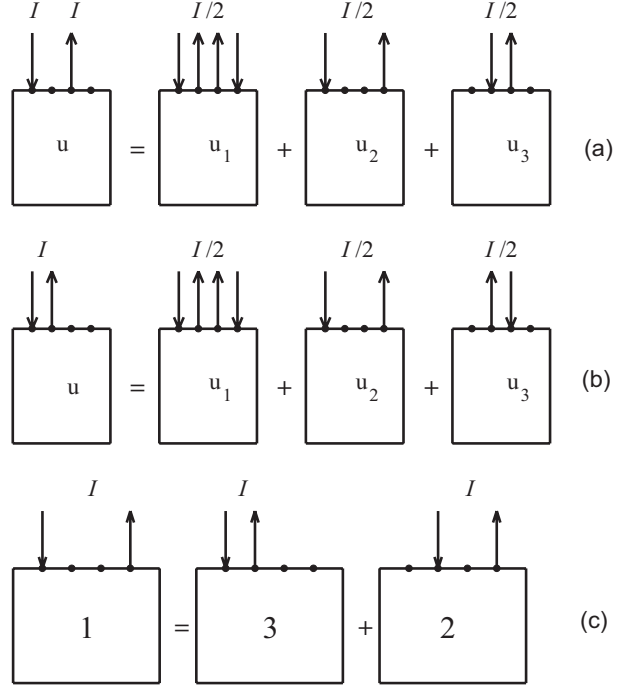}
  \caption{\small  (a) The second collinear geometry as a superposition of three known fields; (b) the third collinear geometry as a superposition of three known fields; and (c) the first collinear geometry as a superposition of the second and third geometries.}
  \label{fig:fig06}
\end{center}
\end{figure}

To improve the convergence, we can use the same technique as that used in Subsection 2.4, i.e., consider that the potential probes are embedded into the sample to the depth $\varepsilon$ and have the coordinates $(\pm s/2, 0, -d/2+\varepsilon)$; after that, we determine the desired value of the function $L_1$, when $\varepsilon$ tends to zero. As a result, we obtain
\begin{equation}\label{eq:2.18}
    \begin{array}{c}
      \displaystyle
      L_1=\frac{16 d^2}{ab}\left[\frac a{2\pi d}\,\frac{\cosh\frac{\pi}a(d-\varepsilon)}{\sinh\frac{\pi d}{a}}\sin\varphi_0\,\sin 3\varphi_0 \right.\\[6mm]
      \displaystyle \left. + {\sum\limits_{k,n=0}^{\infty}}'
      \frac{\theta_n}{\gamma_1d}\,\frac{e^{-\gamma_1\varepsilon}+
      e^{-\gamma_1(2d-\varepsilon)}}{1-e^{-2\gamma_1d}}\,
      \sin\varphi_k\,\sin 3\varphi_k\right],
      \end{array}
\end{equation}
\vspace{0.1mm}

\noindent where $\gamma_1$ and $\varphi_k$ are defined by the formulas

\begin{equation}\label{eq:2.19}
    \begin{array}{c}\displaystyle
      \gamma_1=\pi\sqrt{\left(\frac{2k+1}a\right)^2+\left(\frac{2n}b\right)^2},\\[6mm]
      \displaystyle \varphi_k=\frac{(2k+1)\pi s}{2a}.
    \end{array}
\end{equation}

Let us now consider the second collinear geometry, when a current is fed to the contacts $A$ and $C$ with the coordinates $(-3s/2,0,-d/2)$ and $(+s/2,0,-d/2)$, and the potential difference is measured between the points  $B$ and $D$ with the coordinates $(-s/2,0,-d/2)$ and $(+3s/2,0,-d/2)$. In this case, the desired distribution of the electric field potential in the sample can be obtained as a superposition of three fields in accordance with Fig.~6a, i.e., by the same method as that applied in Subsection 2.4. The field potential $u_1$ is described by formula (\ref{eq:2.8}), the potential $u_2$ is given by formula (\ref{eq:2.3}) at $x_0=3s/2$, and the potential $u_3$ is also described by formula (\ref{eq:2.3}) but at $x_0=s/2$.

After appropriate calculations are performed, the measured potential difference between the contacts $B$ and $D$ and the conditional resistance $R_2=V_{BD}/I_{AC}$ can be calculated. The latter is expressed via the formula $R_2 = ({\rho}/d)L_2$, where the function $L_2$ is expressed by the formula that is analogous to (\ref{eq:2.18}):
\begin{equation}\label{eq:2.20}
    \begin{array}{c}\displaystyle
    L_2=\frac{16 d^2}{ab}\left\{\frac a{2\pi d}\,\frac{\cosh\frac{\pi}a(d-\varepsilon)}{\sinh\frac{\pi d}{a}}\cos^2\varphi_0\,\sin^2 2\varphi_0 \right.\\[6mm]

    \displaystyle -{\sum\limits_{k,n=0}^{\infty}}'\displaystyle\left[
    \frac{\theta_k\theta_n}{\gamma_2d}\frac{e^{-\gamma_2\varepsilon}+e^{-\gamma_2(2d-\varepsilon)}}{1-e^{-2\gamma_2d}}\,
    \sin^2\psi_k\sin^2 2\psi_k  \right. \\[6mm]

    \displaystyle \left. \left. -\frac{\theta_n} {\gamma_1d} \frac{e^{-\gamma_1\varepsilon}+e^{-\gamma_1(2d-\varepsilon)}}{1-e^{-2\gamma_1d}}\cos^2\varphi_k\sin^2 2\varphi_k\right] \right\},
    \end{array}
\end{equation}
where $\gamma_1$ and $\varphi_k$ are defined by formulas (\ref{eq:2.19}), and the parameters $\gamma_2$ and $\psi_k$ have the form
\begin{equation}\label{eq:2.21}
    \gamma_2=\pi\sqrt{\left(\frac{2k}a\right)^2+\left(\frac{2n}b\right)^2}, \qquad \psi_k=\frac{k\pi s}{a}\, .
\end{equation}

The third collinear geometry, where the current contacts are the points $(A,  B)$ and the potential difference is measured between the contacts $(C, D)$, can be considered in a similar manner. The sought field distribution is found as a superposition of three fields in accordance with Fig.~\ref{fig:fig06}b. The potential $u_1$ is described by formula (\ref{eq:2.8}), the potential $u_2$ is given by formula (\ref{eq:2.3}) at $x_0 = 3s/2$ and the current  $I/2$, and the potential $u_3$ is also described by formula (\ref{eq:2.3}) but at $x_0 = s/2$ and the current $-I/2$.

Performing the same calculations, we obtain that the resistance $R_3$ measured in this geometry is expressed by the formula  $R_3 = ({\rho}/d)L_3$, where $L_3$ has the form
\begin{equation}\label{eq:2.22}
    \begin{array}{c}\displaystyle
    L_3=\frac{16 d^2}{ab}\left\{-\frac a{2\pi d}\,\frac{\cosh\frac{\pi}a(d-\varepsilon)}{\sinh\frac{\pi d}{a}}\sin^2\varphi_0\,\cos^2 2\varphi_0 \right. \\[6mm]

    \displaystyle +{\sum\limits_{k,n=0}^{\infty}}'\left[
    \frac{\theta_k\theta_n}{\gamma_2d}\frac{e^{-\gamma_2\varepsilon}+e^{-\gamma_2(2d-\varepsilon)}}{1-e^{-2\gamma_2d}}\,
    \sin^2\psi_k\sin^2 2\psi_k \right. \\[6mm]

    \displaystyle \left. \left. - \frac{\theta_n} {\gamma_1d} \frac{e^{-\gamma_1\varepsilon}+e^{-\gamma_1(2d-\varepsilon)}}{1-e^{-2\gamma_1d}}\sin^2\varphi_k\cos^2 2\varphi_k\right] \right\},
    \end{array}
\end{equation}
\vspace{0.1mm}

Here,  $\gamma_1$, $\gamma_2$, $\varphi_k$, and $\psi_k$ are defined by formulas (\ref{eq:2.19}) and (\ref{eq:2.21}).

The functions $L_1$, $L_2$, and $L_3$ are not independent and are interrelated via the expression $L_1 = L_2 + L_3$, which becomes obvious from Fig.~\ref{fig:fig06}c in which the first collinear geometry is presented as the superposition of the third and second collinear geometries. This can be also demonstrated directly from formulae (\ref{eq:2.18}), (\ref{eq:2.20}), and (\ref{eq:2.22}).

Thus, all the above-described four-probe techniques (collinear arrangement of the probes and the Schnabel method) allow one to obtain the resistivity of an isotropic sample, which has the shape of a rectangular parallelepiped of finite dimensions, from one measurement.\footnote{In all previous sections, point contacts were considered. It is clear that from the experimental standpoint, this imposes appreciable limitations on the current value that passes through the current contacts and, hence, reduces the sensitivity of the method, because the measured potential difference between the potential contacts may be so low that it cannot be measured with the required accuracy. In the study by Schnabel \cite{schnabel67}, an attempt was made to extend the theory of the four-probe technique for measuring the material resistivity to the case of nonpoint contacts, namely, to the case of flat circular contacts that lie on the surface of an infinite conducting space. This problem is of indubitable interest. Unfortunately, the solution that was proposed in \cite{schnabel67} cannot be admitted correct because it was found for a single flat circular contact and cannot be applied to a real problem, in which there are always at least two current contacts of finite dimensions. As a result of a mutual polarization of the contacts, the boundary conditions are not satisfied at their borders. Therefore, the Schnabel solution can be used only in the case where the distance between the contacts far exceeds their dimensions, and the approximation of point contacts then must be used.}

\section{Measurement of the Resistivity of Anisotropic Samples}

Anisotropic conducting substances are of great interest for solid-state physics. It was shown in \cite{vanderpauw61} that the problem of the potential distribution in an anisotropic medium can be reduced to an analogous problem for an isotropic medium by a simple transformation of coordinates.

Let there be a homogeneous conducting anisotropic medium that is characterized by the resistivity tensor $\rho_{ik}$. Let $u(x_1,x_2,x_3)$ be the potential distribution in such a medium. Let us choose the $(x_1,x_2,x_3)$ coordinate system whose axes are directed along the principal axes of the tensor $\rho_{ik}$. In this coordinate system, the tensor $\rho_{ik}$ is diagonal, i.e., it has only three components $(\rho_1,\rho_2,\rho_3)$.

If such a coordinate system is chosen, the equation for the potential $u(x_1,x_2,x_3)$ for an anisotropic medium that follows from the current-continuity equation $(\rm div\:\vec\j=0)$ has the form
\begin{equation}\label{eq:3.1}
    \frac 1{\rho_1}\,\frac{\partial^2u}{\partial x_1^2}+\frac 1{\rho_2}\,\frac{\partial^2u}{\partial x_2^2}+\frac 1{\rho_3}\,\frac{\partial^2u}{\partial x_3^2}=0.
\end{equation}

This equation differs from an ordinary Laplace equation, but using the linear transformation of the coordinate system
\begin{equation*}
    x_i=\beta_i x_i^* \quad(i=1,2,3)
\end{equation*}
it can be easily reduced to standard Laplace equation (\ref{eq:2.1}) for an isotropic conducting medium with a certain resistivity $\rho^*$. In this case, $u(x_1,x_2,x_3)$ transforms into $u^*(x_1^*,x_2^*,x_3^*)$, i.e., the potential values at the corresponding points of the initial anisotropic medium and its isotropic image are identical. Hereinafter, the asterisk in the index always designates the parameter of the isotropic medium–image, which corresponds to the same parameter of its anisotropic prototype.

The coefficients $\beta_1,\,\beta_2,\,\beta_3$, and the resistivity $\rho^*$ of the isotropic medium are chosen according to \cite{vanderpauw61} so that the currents through any cross sections in the initial anisotropic system are equal to the currents through the corresponding cross sections of its isotropic image, namely:
\begin{equation}\label{eq:3.2}
    \rho^*=(\rho_1 \rho_2 \rho_3)^{1/3}\, ,
\end{equation}
and
\begin{equation*}
    \beta_1=\left(\frac{\rho^*}{\rho_1}\right)^{1/2},\ \beta_2=\left(\frac{\rho^*}{\rho_2}\right)^{1/2},\ \beta_3=\left(\frac{\rho^*}{\rho_3}\right)^{1/2}.
\end{equation*}

This means that the conditional resistances measured as the ratio of the potential difference between the potential contacts to the value of the current that flows between the current contacts in the initial anisotropic sample and its isotropic image are identical, thus allowing application of the results obtained for isotropic samples to anisotropic samples.

\subsection{Four Collinear Contacts}

It was shown in Subsection 2.5 that for an isotropic sample of finite dimensions, the conditional resistances $R_1$ and $R_3$ that are experimentally measured in the first and third collinear geometries are expressed by the formulas
\begin{equation*}
    R_1=\frac{\rho}{d}\,L_1(a,b,d,s),\quad R_3=\frac{\rho}{d}\,L_3(a,b,d,s),
\end{equation*}
where the functions $L_1$ and $L_3$ depend only on the geometrical parameters of the experiment, i.e., on the four dimensions $a,b,d$, and $s$, and are expressed by formulas (\ref{eq:2.18}) and (\ref{eq:2.22}). In reality, as is seen from these formulas, $L_1$ and $L_3$ are functions of only three independent arguments for which the ratios $a/s,\;b/s$, and $\alpha=d/s$ can be taken.

For an anisotropic sample, the measured resistances $R_1$ and $R_3$, as was mentioned above, are the same as $R_1^*$ and $R_3^*$ for the isotropic image with the resistivity $\rho^*$, which is determined from formula (\ref{eq:3.2}), and dimensions $a^*=(\rho_1/\rho^*)^{1/2}a,\;b^*=(\rho_2/\rho^*)^{1/2}b$, and $d^*=(\rho_3/\rho^*)^{1/2}d$. The distance between the contacts in the isotropic model is $s^*=(\rho_1/\rho^*)^{1/2}s$, if they are arranged along the $x_1$ axis in the actual anisotropic sample. Hence, the expressions for  $R_1$ and $R_3$ for the anisotropic sample take the form
\begin{equation}\label{eq:3.3}
    \begin{array}{c}
    \displaystyle R_1=\frac{\sqrt{\rho_1\rho_2}}{d}\,L_1(a^*/s^*,b^*/s^*,\alpha^*),\\[6mm] \displaystyle R_3=\frac{\sqrt{\rho_1\rho_2}}{d}\,L_3(a^*/s^*,b^*/s^*,\alpha^*),
    \end{array}
\end{equation}
where $a^*/s^*=a/s,\;b^*/s^*=(\rho_2/\rho_1)^{1/2}b/s$, and $\alpha^*=(\rho_3/\rho_1)^{1/2}\alpha$, because the contacts are positioned along the $x_1$ axis. If, in addition, the condition $\rho_1=\rho_2$ (no anisotropy of $\rho$ in the $(x_1, x_2)$ plane is present) is met, the experimentally measured ratio of the resistances assumes the form
\begin{equation}\label{eq:3.4}
    \frac{R_1}{R_3}=\frac{R_1^*}{R_3^*}=\frac{L_1(a/s,b/s,\alpha^*)}{L_3(a/s,b/s,\alpha^*)}=
    {F_{13}(a/s,b/s,\alpha^*)}.
\end{equation}

\begin{figure}[h]
\begin{center}
\includegraphics[width=85mm]{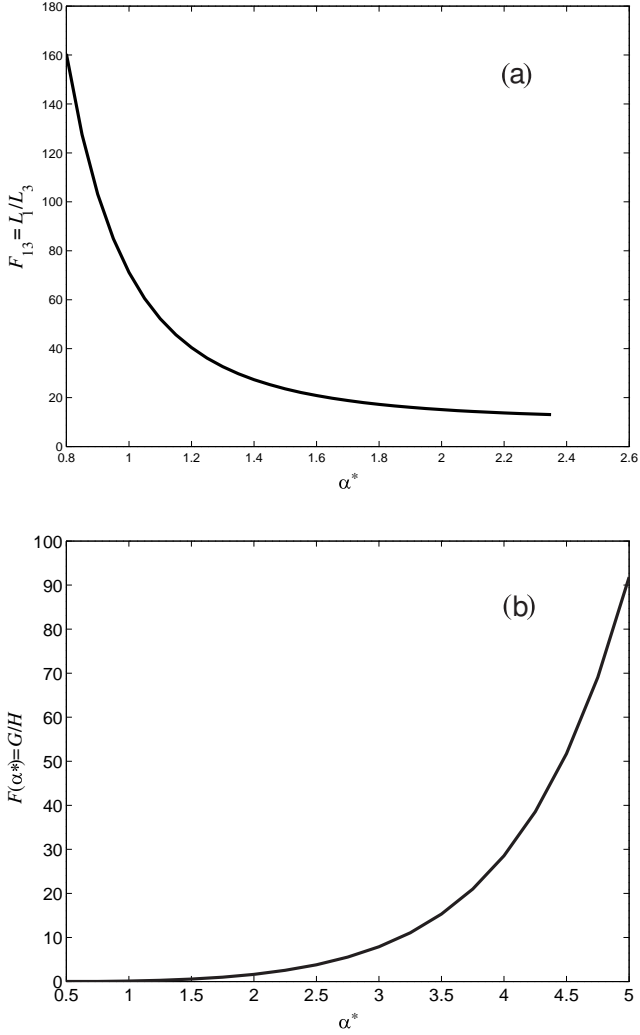}
  \caption{\small Plots of the functions  $F_{13}(\alpha^*)=L_1(\alpha^*)/L_3(\alpha^*) \ (a) $
  and $F(\alpha^*)=G(\alpha^*)/H(\alpha^*)$ (b) for a sample with the
  dimensions $a = 9$~mm, $b = 3$~mm, and $s = 2.5$~mm.}
  \label{fig:fig07}
\end{center}
\end{figure}

If the ratios $a/s$ and $b/s$ are considered as parameters and $\alpha^*$ as an argument, the function $F_{13}$ can be plotted as a function of the argument $\alpha^*$. As an example, Fig.~\ref{fig:fig07}a shows the plot of $F_{13}$ calculated from formula (\ref{eq:3.4}) using formulas (\ref{eq:2.18}) and (\ref{eq:2.22}) for a sample with dimensions $a = 9$~mm, $b = 3$~mm, and $s = 2.5$~mm.

Thus, by plotting the function $F_{13}(a/s,b/s,\alpha^*)=F_{13}(\alpha^*)$ for a particular sample with the known parameters $a/s$ and $b/s$, we can find from this plot, using the measured value of the ratio $R_1/R_3$, the value of the argument $\alpha^*=(\rho_3/\rho_1)^{1/2}\alpha$ that corresponds to it and, hence, determine the degree of anisotropy of $\rho$ along the $x_3$ and $x_1$ axes using the known geometrical parameters $d$ and $s$:
\begin{equation}\label{eq:3.5}
    \frac{\rho_3}{\rho_1}=\left(\frac{\alpha^*}{\alpha}\right)^2=\left(\frac sd\,\alpha^*\right)^2.
\end{equation}

The resistivity $\rho_1$ can now be found from any formula (\ref{eq:3.3}):
$$
\rho_1=\frac{R_1d}{L_1(a/s,b/s,\alpha^*)}=\frac{R_3d}{L_3(a/s,b/s,\alpha^*)},
$$
where the functions $L_1$ and $L_3$ are calculated from formulae (\ref{eq:2.18}) and (\ref{eq:2.22}), and the resistivity $\rho_3$ is then found from (\ref{eq:3.5}).

\subsection{Schnabel Method}

Measurements according to the Schnabel method \cite{schnabel64,schnabel67} should be performed in the same way as the procedure described above. In previous sections, expressions (\ref{eq:2.13}) and (\ref{eq:2.10}) were presented for the conditional resistances $R_1$ and $R_2$, which are measured for an isotropic sample in two Schnabel geometries. In these formulas, $R_1$ and $R_2$ are expressed using calculated functions $G$ (\ref{eq:2.15}) and $H$ (\ref{eq:2.11}), which depend only on the geometry of the experiment. As in Subsection 3.1, the ratio of the measured resistances $R_1/R_2$ is a function of only three independent arguments: $a^*/s^*, b^*/s^*$, and $a^* =  d^*/s^*$, where $a^*/s^* = a/s, \ b^*/s^* = (\rho_2/\rho_1)^{1/2}b/s$, and $\alpha^*=(\rho_3/\rho_1)^{1/2}\alpha$, if the contacts are positioned along the $x_1$ axis.

If the condition $\rho_1=\rho_2$ is satisfied (no anisotropy of $\rho$ in the $(x_1, x_2)$ plane is observed), the experimentally measured ratio of resistances assumes the form
\begin{equation}\label{eq:3.6}
    \frac{R_1}{R_2}=\frac{R_1^*}{R_2^*}=F(a/s, b/s, \alpha^*)=
    \frac{G(a/s, b/s, \alpha^*)}{H(a/s, b/s, \alpha^*)}.
\end{equation}

Thus, by plotting the function $F(a/s, b/s, \alpha^*)= F(\alpha^*)$ for a particular sample with the known parameters $a/s$ and $b/s$ we can find the value of the argument $\alpha^*=(\rho_3/\rho_1)^{1/2}\alpha$ from this plot using the measured ratio of the resistances $R_1/R_2$ corresponding to this argument. After that, using the same procedure as in Subsection 3.1, the anisotropy $\rho_3/\rho_1$ is found from formula (\ref{eq:3.5}), and the resistivity $\rho_1$ is determined as
\begin{equation}\label{eq:3.7}
\rho_1=\frac{R_1d}{G(a/s,b/s,\alpha^*)}=\frac{R_2d}{H(a/s,b/s,\alpha^*)}.
\end{equation}

Figure~\ref{fig:fig07}b shows the plot of the function $F(\alpha^*)=R_1/R_2$ constructed according to formula (\ref{eq:3.6}) using formulae (\ref{eq:2.15}) and (\ref{eq:2.11}) for a sample with dimensions $a = 9$~mm, $b = 3$~mm, and $s = 2.5$~mm.

\section{Measuring the Temperature Dependence of the Resistivity of Highly Oriented Pyrolitic Graphite}

The above-described technique was used in practice to measure the resistivity of such a highly anisotropic material as highly oriented pyrolitic graphite (HOPG) within a temperature range  $T$ = 4.2--293~K using the Schnabel method. A massive HOPG sample was manufactured at the OAO NIIGrafit and, according to the data of X-ray diffraction investigations, had an angular disorientation of crystallites of $\sim$~0.3$^\circ$. In our experiment, the HOPG sample had the following dimensions: (length) $a = 9$~mm, (width) $b = 3$~mm, and (thickness) $d = $0.03~mm; the distance between the contacts was $s = 2.5$~mm. The plot of the function $F(\alpha^*)=G(\alpha^*)/H(\alpha^*)$ in Fig.~\ref{fig:fig07}b was presented, as an example, just for these dimensions. The sample was fixed in a special holder, and contacts to the sample were manufactured from a 30$\mu$m-diameter copper wire and attached to it with a self-solidifying silver paste. The experiments were performed in a blown through Dewar--insert for intermediate temperatures. The temperature stabilization and sweep were performed with a LakeShore model 331 temperature controller.

\begin{figure}[h]
\begin{center}
\includegraphics[width=80mm]{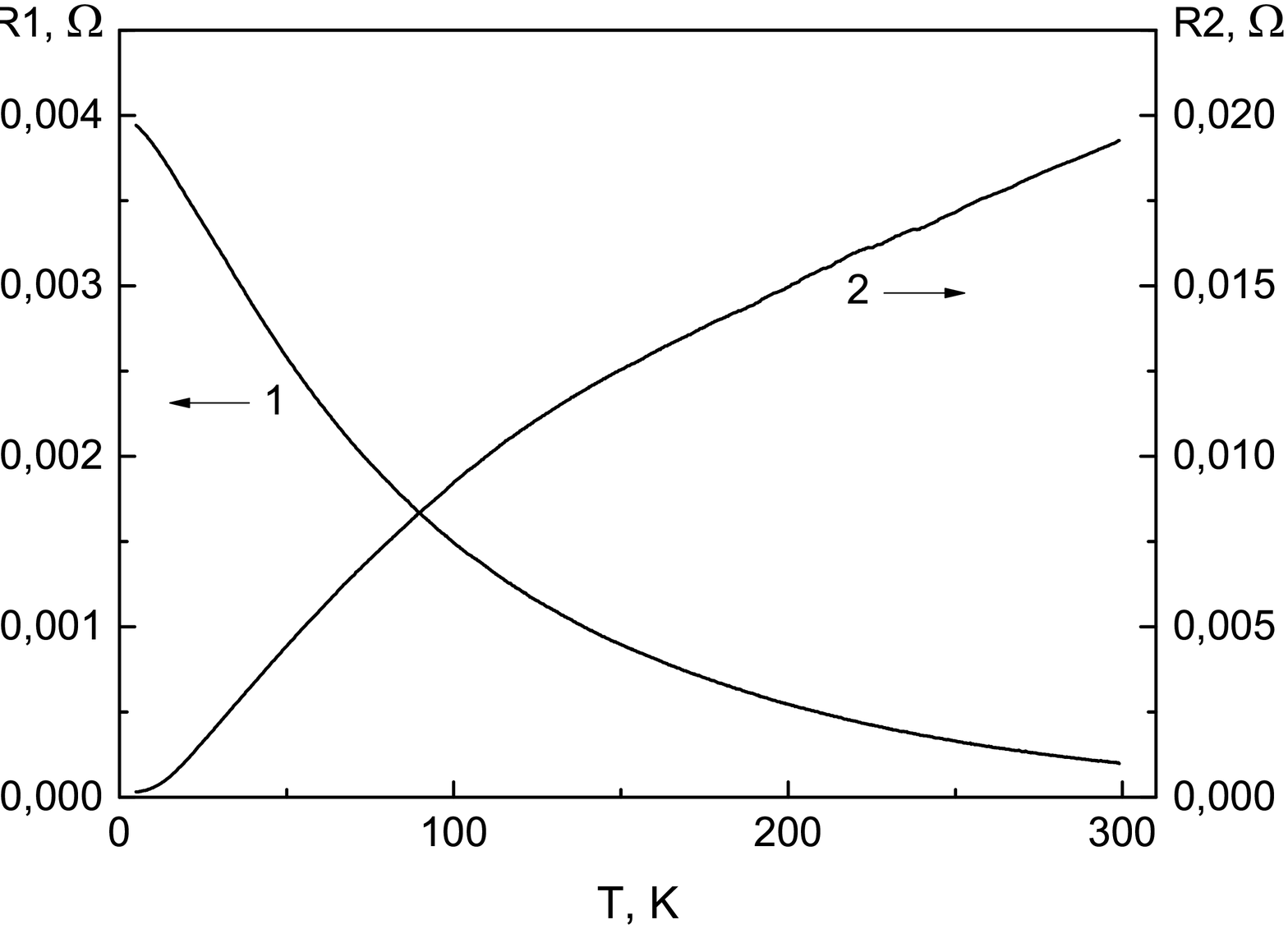}
  \caption{\small Experimental temperature dependences of the conditional resistances $R_1$ (1) and $R_2$ (2) in the first and second Schnabel geometries for a 30$\mu$m-thick HOPG sample.}
  \label{fig:fig08}
\end{center}
\end{figure}

First, the temperature dependences of the conditional resistances $R_1$ and  $R_2$ were measured. The obtained experimental data (Fig.~\ref{fig:fig08}) for the further processing were represented in the form of a data array of three columns $(T,R_1,R_2)$, where $T$ is the sample temperature, and $R_1$ and $R_2$ are the conditional resistance that were measured in the first and second Schnabel geometries at this temperature. The number of rows $n$ in the array was determined by the number of experimental temperature points (in our case, $n = 1180$ at a temperature step of 0.25~K).

The experimental data were processed in two stages. At the first stage, the functions  $G(\alpha^*)$, $H(\alpha^*)$, and $F(\alpha^*)=G(\alpha^*)/H(\alpha^*)$ were calculated from the known parameters $(a,b,s)$ using formulas (\ref{eq:2.11}) and (\ref{eq:2.15}). The range of values of the argument $\alpha^*$ was determined by the minimum and maximum values of the ratio $R_1/R_2$ from the experimental data array $(T,R_1,R_2)$. At this stage, it is necessary to correctly specify the depth values of the potential contacts $\varepsilon_1$ and $\varepsilon_2=2\varepsilon_1$ and the number of diagonals $N$ along which the series are summed for the functions $G$ and $H$ in order to obtain the calculation results with the required accuracy. At low $\alpha^*$ values, the number of diagonals $N$ may be rather large (about several thousands). As the argument $\alpha^*$ increases, the number of required diagonals $N$ appreciably decreases. For example, the plot in Fig.~\ref{fig:fig07}b is constructed in the range $\alpha^*$ = 0.5--5 with a step of 0.25 using 19 points at $\varepsilon_1=0.01d$ and $\varepsilon_2=2\varepsilon_1$. In this case, the number of considered diagonals along which summing is performed in the range $\alpha^*$ = 0.5--0.8 is $N$ = 1600, in the range $\alpha^*$ = 0.8--1, $N$ = 1000, and at $\alpha^*>1$ –-- $N=600$. The calculation result at this stage is an array of numbers that contains four columns $(\alpha^*,G,H,F)$, and the number of rows in this array is determined by the range of values of the argument $\alpha^*$ and the chosen step in this argument.

At the second stage, the experimental data are actually processed. The processing program approximates the curves $G(\alpha^*)$ and $H(\alpha^*)$, which were obtained at the first stage, by cubic splines and constructs the function $\alpha^*=F^{-1}(R_1/R_2)$, which is also approximated by cubic splines. The processing program sequentially chooses points from the experimental data array $(T,R_1,R_2)$, finds the ratio $R_1/R_2$, and then determines the value of $\alpha^*=F^{-1}(R_1/R_2)$ from the approximating formula. The anisotropy is determined from formula (\ref{eq:3.5}) using the found value of $\alpha^*$. Subsequently, two values, $\rho_{1g}$ and $\rho_{1h}$, are found for the resistivity $\rho_1$ from formula (\ref{eq:3.7}), into which the values of the functions $G(\alpha^*)$ and $H(\alpha^*)$ are substituted. These values are obtained using the above-determined approximating formulas. The arithmetic mean of the $\rho_{1g}$ and $\rho_{1h}$ values is taken as $\rho_1$, i.e., $\rho_1=(\rho_{1g}+\rho_{1h})/2$, and the value of $\rho_3$ is then found from formula (\ref{eq:3.5}).

Figure 9 shows the results of such processing in the form of the temperature dependences of $\rho_1$ and $\rho_3$ for the investigated HOPG sample and the temperature dependence of the value of the resistivity anisotropy $\rho_3/\rho_1$.

\begin{figure}[h]
\begin{center}
\includegraphics[width=88mm]{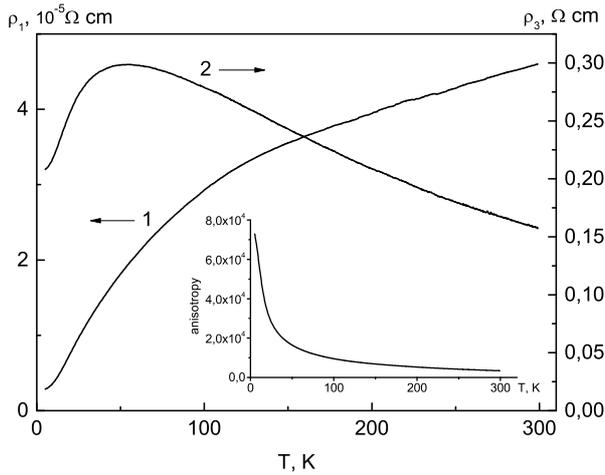}
  \caption{\small Temperature dependences of the resistivities $\rho_1$ (1) and  $\rho_3$ (2) for the HOPG sample (obtained via processing of the data presented in Fig.~\ref{fig:fig08}). The insert shows the temperature dependence of the resistivity anisotropy $\rho_3/\rho_1$.}
  \label{fig:fig09}
\end{center}
\end{figure}

\section{Conclusion}

Hence, both the Schnabel method and the method of four collinear contacts include two independent measurements of conditional resistances, which do not provide the determination of all three components of the resistivity tensor. However, if the anisotropy of a studied substance is such that the condition  $\rho_1=\rho_2$ is fulfilled, both methods allow one to find the degree of the anisotropy $(\rho_3/\rho_1)$ and then both values of $\rho_1$ and $\rho_3$. The condition $\rho_1=\rho_2$ is satisfied for most substances that are interesting for practice. Depending on the degree of crystal anisotropy and its dimensions, one of these methods should be chosen in order to determine the desired parameters of a sample with the maximum accuracy.

For example, for a sample with the dimensions $a = 9$~mm, $b = 3$~mm, and $s = 2.5$~mm, it follows from Figs.~\ref{fig:fig07}a and 7b that for ``thin'' samples $(a^* < 1.5)$, a collinear arrangement of contacts provides a higher accuracy, while the Schnabel method should be preferred for ``thick'' samples $(a^* > 2)$.

For materials in which all three resistivity components are different, other arrangements of contacts must be used. There are various possibilities: other orientations of the line of contacts, placing the potential contacts on the sample sides that are adjacent to sides on which the current contacts are positioned, etc.

For samples of finite dimensions that are shaped as rectangular parallelepipeds, the above-described methods make it possible to successfully find the electric field potential distribution inside a sample and solve the formulated problem. In this case, the calculations are rather simple and accurate, thus allowing one not to use approximations in which one or two sample dimensions are considered infinite and consider an actual sample from the very beginning. Anyway, the technique described in this study allows determination of errors that arise when using the approximations in which one, two, or all the three dimensions of a sample are assumed to be infinite.

\section*{Acknowledgments}

We are very grateful to I.A.~Fomin for useful discussions, A.A.~Dolgoborodov for his help in performing measurements of the resistivity of highly oriented pyrolitic graphite, and A.S.~Kotosonov for an HOPG sample that was kindly given for measurements.

\end{document}